\documentclass[twoside,superscriptaddress,twocolumn,a4paper,floatfix,prl,aps,longbibliography]{revtex4-1}

\usepackage{xcolor}
\usepackage{graphicx}
\usepackage[utf8x]{inputenc}
\usepackage[T1]{fontenc}
\usepackage{siunitx}
\usepackage{amstext}
\usepackage{hyperref}
\usepackage{amsfonts}
\usepackage{amsmath}
\usepackage{tabularx}
\usepackage{filecontents}
\usepackage{soul}
\usepackage{dsfont}
\usepackage{mathtools}
\usepackage{makecell}
\usepackage{fourier, heuristica}
\usepackage{multirow}
\usepackage{bbold}
\usepackage{fdsymbol}
\usepackage{wasysym}

\def\block(#1,#2)#3{\multicolumn{#2}{c}{\multirow{#1}{*}{$ #3 $}}}

\begin{document}

\title {How confused can an entanglement witness be to be still persuasive}

\author{Jan Roik}
\email{jan.roik@upol.cz}
\affiliation{RCPTM, Joint Laboratory of Optics of Palacký University and Institute of Physics of Czech Academy of Sciences, 17. listopadu 12, 771 46 Olomouc, Czech Republic}

\author{Karol Bartkiewicz} \email{bartkiewicz@jointlab.upol.cz}
\affiliation{Faculty of Physics, Adam Mickiewicz University,
PL-61-614 Pozna\'n, Poland}
\affiliation{RCPTM, Joint Laboratory of Optics of Palacký University and Institute of Physics of Czech Academy of Sciences, 17. listopadu 12, 771 46 Olomouc, Czech Republic}

\author{Antonín Černoch} \email{antonin.cernoch@upol.cz}
\affiliation{RCPTM, Joint Laboratory of Optics of Palacký University and Institute of Physics of Czech Academy of Sciences, 17. listopadu 12, 771 46 Olomouc, Czech Republic}
   
\author{Karel Lemr}
\email{k.lemr@upol.cz}
\affiliation{RCPTM, Joint Laboratory of Optics of Palacký University and Institute of Physics of Czech Academy of Sciences, 17. listopadu 12, 771 46 Olomouc, Czech Republic}

\begin{abstract}
Detection of entangled states is essential in both fundamental and applied quantum physics. However, this task proves to be challenging especially for general quantum states. One can execute full state tomography but this method is time demanding especially in complex systems. Other approaches use entanglement witnesses, these methods tend to be less demanding but lack reliability. Here, we demonstrate that ANN -- artificial neural networks provide a balance between both approaches. In this paper, we make a comparison of ANN performance against witness-based methods for random general 2-qubit quantum states without any prior information on the states. Furthermore, we apply our approach to real experimental data set.  
\end{abstract}

\date{\today}

\maketitle
\paragraph*{Introduction.} Quantum entanglement is an intriguing phenomenon described almost a century ago by Schrödinger, Einstein, Podolsky, and Rosen \cite{schrodinger_1935,PhysRev.47.777}. Since then many theoretical and practical papers alike, as well as vivid discussions, were dedicated to this topic \cite{plenio2014introduction,RevModPhys.81.865,MINTERT2005207}. The ability to effectively detect entangled state became essential mainly because of their application potential in quantum computing \cite{Steane_1998}, quantum cryptography \cite{RevModPhys.74.145}, and quantum teleportation experiments \cite{Bouwmeester1997}. The most robust way of detecting it is via a full state tomography and density matrix estimation \cite{PhysRevA.78.022322}. This method allows us to obtain all information about the state and thus correctly detect  entanglement. Unfortunately this method is experimentally demanding because the number of required projections grows exponentially with the dimension of Hilbert space. There is also a variety of other methods that do not rely on full-state tomography \cite{PhysRevLett.23.880,PhysRevA.88.052105,PhysRevA.68.052101,PhysRevLett.95.240407,PhysRevLett.93.230501,PhysRevLett.109.200503,PhysRevLett.97.050501,Walborn2006,PhysRevLett.100.140403,PhysRevLett.104.210501,PhysRevLett.98.110502,Eisert_2007,PhysRevA.77.030301,PhysRevA.81.022307,PhysRevLett.107.150502,PhysRevLett.106.190502,PhysRevA.86.062329,Rudnicki_2014,PhysRevA.90.024301,PhysRevLett.105.230404,PhysRevLett.108.240501}. These methods include a wide range of linear entanglement witnesses \cite{PhysRevA.88.052105,PhysRevA.68.052101,PhysRevLett.95.240407,PhysRevLett.93.230501,PhysRevLett.109.200503,PhysRevLett.97.050501} of the CHSH -- Clauser Horne Shimony Holt type \cite{PhysRevLett.23.880}. 
While for pure states, these methods give similar results, their outcomes might vary significantly when mixed states are considered. While requiring only a relatively few measurement configurations, these witnesses can not reliably function without some a prior information about the detected state. To circumvent this limitation, while not resorting to state tomography, non-linear entanglement witnesses have been proposed. 
\begin{figure}[ht!]
		\begin{center}
		\includegraphics[scale=0.2]{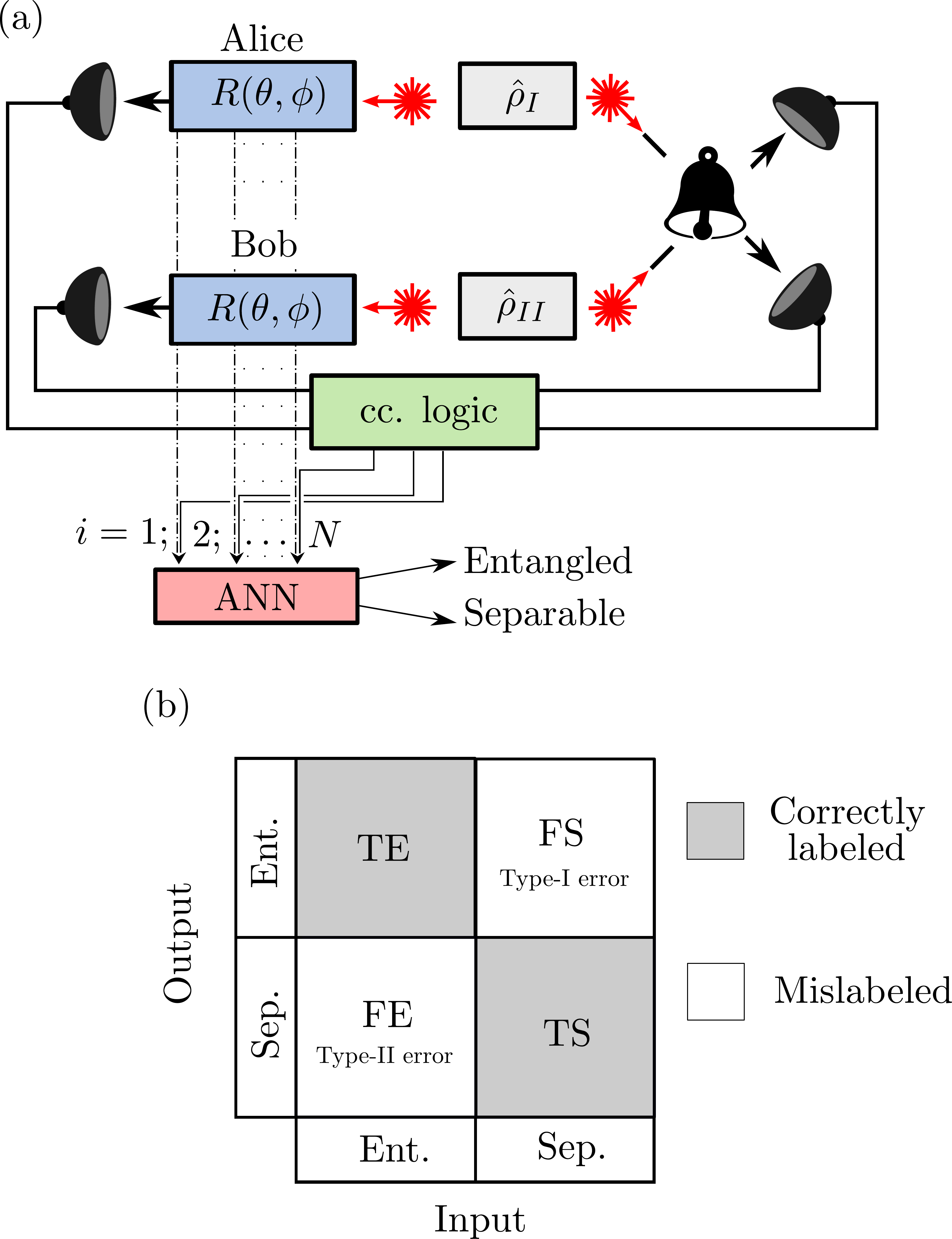}
		\caption{(a) Scheme of collectibility measurement:  Two copies $\hat{\rho}_{I},\hat{\rho}_{II}$ of the same quantum state are generated. One qubit from each pair is measured locally meanwhile remaining qubits are subjected to Bell measurement. Collectibility is then calculated analytically from obtained coincidence detections. Alternatively, as we investigate in this paper, coincidence detections can be fed to an ANN (artificial neural network) which then labels the states. (b) Schematic depiction of the confusion matrix used for performance evaluation of the ANN. TE -- truly entangled, FE -- falsely entangled, TS -- truly separable, FS -- falsely separable, Sep. -- separable, Ent. -- entangled.}
		\label{fig:scheme}		
		\end{center}
\end{figure}

In 2011, Rudnicki \textit{et al.} introduced a nonlinear entanglement witness called \textit{Collectibility} \cite{PhysRevLett.107.150502,PhysRevA.86.062329}. For a visual demonstration of this concept [see Fig.\ref{fig:scheme} (a)]. For 2-qubit states, this witness requires two simultaneously prepared copies of the investigated state. Then a Bell state projection is imposed on a pair of corresponding qubits from each copy and the remaining qubits are subjected to local measurements. For a general 2-qubit state, this requires a combination of 5 local projections and, thus, fewer measurement configuration than full quantum state tomography which includes at least $24$ projections. One can further decrease the time needed for a QST if measurements can be performed in parallel on multiple copies of the investigated state. When dealing with unknown quantum states, collectibility can detect a much broader range of states compared to linear witnesses. Namely, it detects all pure entangled states. Unfortunately, it detects entanglement of only a fraction of mixed states. This shortcoming is characterized by a rather big Type-II error (false negative), as we show later. On the other hand, all states which are classified as entangled by this method are classified correctly (Type-I error is null, there are no false-positive classifications). We demonstrate that significant improvement can be reached when collective entanglement witnesses are devised using an artificial neural network. As demonstrated by Gao \textit{et al.} \cite{PhysRevLett.120.240501} and other groups \cite{Ma2018,PhysRevA.98.012315}, neural networks can be used to identify quantum states. However, only linear entanglement witnesses were considered which significantly limited the class of detected entangled states. Note that neural network-based linear witnesses share the same shortcomings with their analytical counterparts, which is the need for a prior information about the investigated state. 

We train a neural network to classify quantum states by providing it with results of collective measurements and demonstrate its significantly better performance over collectibility and other similar non-linear witnesses for a general 2-qubit state as well as for real experimental data for a fixed number of measurement configurations. Moreover, we show the increasing capability of the neural network when provided with a larger amount of measurement configuration outcomes by comparing it against three other analytical methods that require 12 projections, namely FEF -- fully entangled fraction \cite{PhysRevA.54.3824,PhysRevA.95.030102,PhysRevA.91.032315}, CHSH \cite{PhysRevLett.23.880}, and entropic witness \cite{Castro_2003,PhysRevA.57.1619}. These projections are listed in \cite{Suplement}. We use confusion matrix as a method of performance evaluation for the ANN and previously known non-linear witnesses [see Fig.\ref{fig:scheme} (b)]. Diagonal elements show the number of correctly labeled input states TE -- truly entangled and TS -- truly separable furthermore off-diagonal elements provide information about falsely labeled input states FE -- falsely entangled and FS -- falsely separable.

\paragraph*{Neural network.}Random two-qubit states $\hat{\rho}$ were generated \cite{Suplement}. The state of two of its copies is described by density matrix $\hat{\rho}_T = \hat{\rho} \otimes \hat{\rho}_n$ where $\hat{\rho}_n$ is derived from $\hat{\rho}$ by swapping subsystems. Subsequently, $\hat{\rho}_T$ was subjected to projective measurements and probabilities were obtained
\begin{equation}
P_{xy}= \frac{T_r[\hat{\rho}_t\hat{\pi}_x\hat{\rho}_b\hat{\pi}_y]}{T_r[\hat{\rho}_t\hat{\pi}_x\  \hat{\mathbb{1}}\hat{\pi}_y]},
\end{equation}
where $\hat{\pi}_x$ and $\hat{\pi}_y$ are local projections onto single-qubit states $|x\rangle$ and $|y\rangle$ and $\hat{\rho}_B$ is the density matrix of the singlet Bell state. An obtained set of $N$ probabilities $P_{xy}^{(i)}\; ; i = 1,...,N$ is subsequently fed to a neural network for training together with labels obtained by the PPT- Peres-Horodecki criterion \cite{HORODECKI19961,PhysRevLett.77.1413}.

\begin{figure}
		\begin{center}
		\includegraphics[scale=0.17]{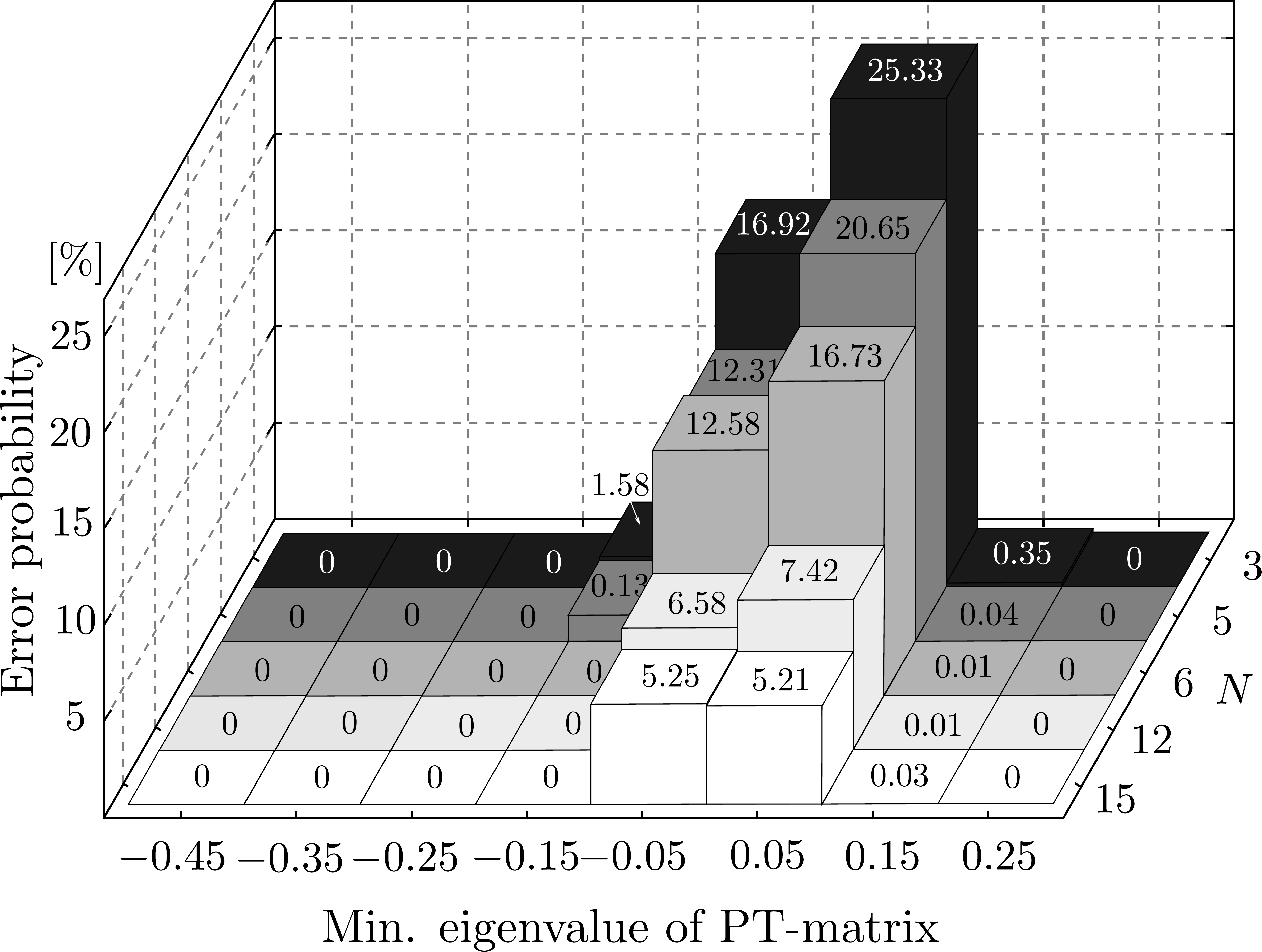}
		\caption{The result obtained by the neural network with decision threshold $\epsilon= 0.5$ for $N=3,5,6,12,15$ and distribution containing $67.74$ $\%$ entangled states and $32.26$ $\%$ separable states. In this graph probability of false prediction is plotted against the minimal eigenvalue of PT-matrix.}
		\label{fig:original_trashold}		
		\end{center}
\end{figure}

TensorFlow 2.0 \cite{abadi2016tensorflow} was used to program a neural network capable of classifying quantum states. We experimented with the complexity of the network and our final layout of the network with $5$ hidden layers containing $36, 180, 75, 180,$ and $75$ nodes respectively seems to be the optimal choice to find a balance between obtained precision and computation time. The proposed network is capable of assigning any quantum state with a value $w \in [0;1]$ which can be interpreted as a confidence factor from 0 (certainly entangled) to 1 (certainly separable). We defined decision threshold $\epsilon$ to convert the $w$ values to a binary label: $w<\epsilon \Rightarrow$ entangled,  $w\geq \epsilon \Rightarrow$ separable. By changing $\epsilon$ value we make the network biased towards the desired decision which allowed us to tune the trade-off between Type-I and Type-II errors. The network was trained on $4\times 10^6$ samples and tested on the other $4\times 10^5$ samples with distribution containing $67.74$ $\%$ entangled states and $32.26$ $\%$ separable states. For more details about the purity distribution of the samples see \cite{Suplement}. The main goal was to test the network against collectibility, therefore, we start to train it using the same $N=5$ projection settings (see Supplemental material \cite{Suplement} for a brief overview on collectibility). In the next step, we also tested capability of the network for $N= 3,6,12,15$ projection settings (see \cite{Suplement} for more details).
\paragraph*{Results.} In the first step, we decided to test the neural network with decision threshold $\epsilon= 0.5$ for a several amounts of projection settings $N=3,5,6,12,15$. As it turns out the neural network was capable of labeling entangled and separable states even using 3 projection settings with an overall success rate of around $83.33$ $\%$. For an increasing number of projection settings success rate increased even further and reached $96.55$ $\%$ for $15$ projections settings. We plot the probability of incorrect decision as a function of the smallest eigenvalue of the partially transposed density matrix $\hat{\rho}$ (see Fig. \ref{fig:original_trashold}). 
\begin{figure}
		\begin{center}
		\includegraphics[scale=0.17]{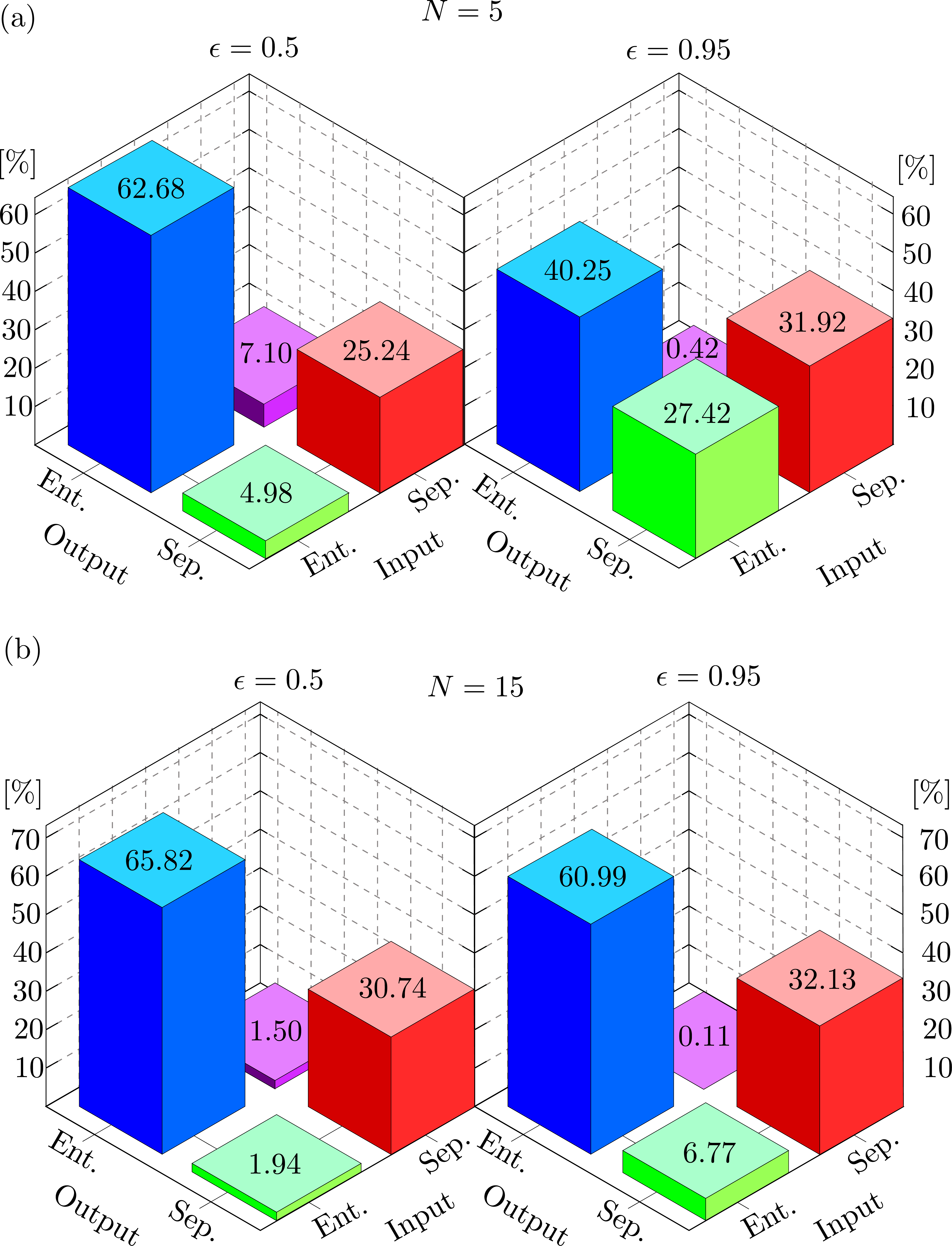}
		\caption{Performance dependence of the ANN on decision threshold $\epsilon= 0.5,0.95$ with distribution containing $67.74$ $\%$ entangled states and $32.26$ $\%$ separable states depicted as confusion matrices for: (a) $N =5$ ; (b) $N=15$.}
		\label{fig:CM_N5}		
		\end{center}
\end{figure}
As expected, the neural network struggles with the states close to the PPT decision boundary (minimal eigenvalue close to zero). Unfortunately, the neural network is, to some extend, prone to Type-I errors (separable state classified as entangled). As it turns out the network is more likely to make a mistake when classifying separable states than entangled states. Our solution is to change the decision threshold $\epsilon$ to decrease the Type-I error. This means that we demand more certainty from the network when classifying the entangled state. By optimizing thresholds we manage to find the value which satisfies a condition of Type-I error $<1$ $\%$ which we find acceptable. It is possible to arbitrarily decrease the Type-I error by sacrificing the detection capability characterized by Type-II error. For more detailed dependence of Type-I and Type-II error on threshold  for $N=3,5,6,12,15$ see Fig. \ref{fig:CM_N5} and \cite{Suplement}. In the next step we compared the network performance against collectibility. The neural network fed by outcomes of the same 5 projection settings also required by the collectibility was able to correctly classify $78.14$ $\%$ of all states while committing Type-I error of $0.96$ $\%$ $(\epsilon = 0.9)$. This performance vastly surpassed the capability of the Collectibility which identifies only $36.59$ $\%$ of the states correctly (see Tab. \ref{Tab:trashold}). To further highlight the potential of ANN we compared its performance with analytical methods (FEF, CHSH, and EW) (see Tab. \ref{Tab:trashold}). The success rate of the ANN surpass capabilities of FEF by $6.01$ $\%$, EW by $34.01$ $\%$, and CHSH by $46.01$ $\%$ while committing Type-I error $0.24$ $\%$. This means that if we can accept some Type-I error, it is possible to achieve a major improvement in entangled states detection using the neural network.

\begin{table}[]
\begin{tabular}{cccccc}
\hline
                      & \multicolumn{5}{c}{ANN}                       \\ \hline
N                     & 3             & 5     & 6     & 12    & 15    \\ \hline
Type-I error {[}\%{]} & 0.93          & 0.96  & 1.18  & 0.24  & 0.22  \\
Type-II error {[}\%{]} & 33.47         & 20.91 & 15.88 & 7.74  & 5.24  \\
Success rate {[}\%{]} & 65.50         & 78.14 & 82.94 & 92.01 & 94.54 \\ \hline
                      &               &       &       &       &       \\ \hline
                      & Colectibility & FEF   & EW    & CHSH  &       \\ \hline
N                     & 5             & 12    & 12    & 12    &       \\ \hline
Type-I error {[}\%{]} & 0             & 0     & 0     & 0     &       \\
Type-II error {[}\%{]} & 63.41         & 14.00 & 42.00 & 54.00 &       \\
Success rate {[}\%{]} & 36.59         & 86.00 & 58.00 & 46.00 &       \\ \hline
\end{tabular}
\caption{Comparison of the results obtained by ANN for  $N= 3,5,6,12,15$ with prominent analytical methods (collectibility, FEF -- fully entangled fraction, EW -- entropic witness, CHSH nonlocality).  Both Type-I and Type-II errors are taken for decision threshold $\epsilon = 0.9$ to ensure Type-I error $< 1\%$.}
\label{Tab:trashold}
\end{table}

\begin{figure}
		\begin{center}
		\includegraphics[scale=0.21]{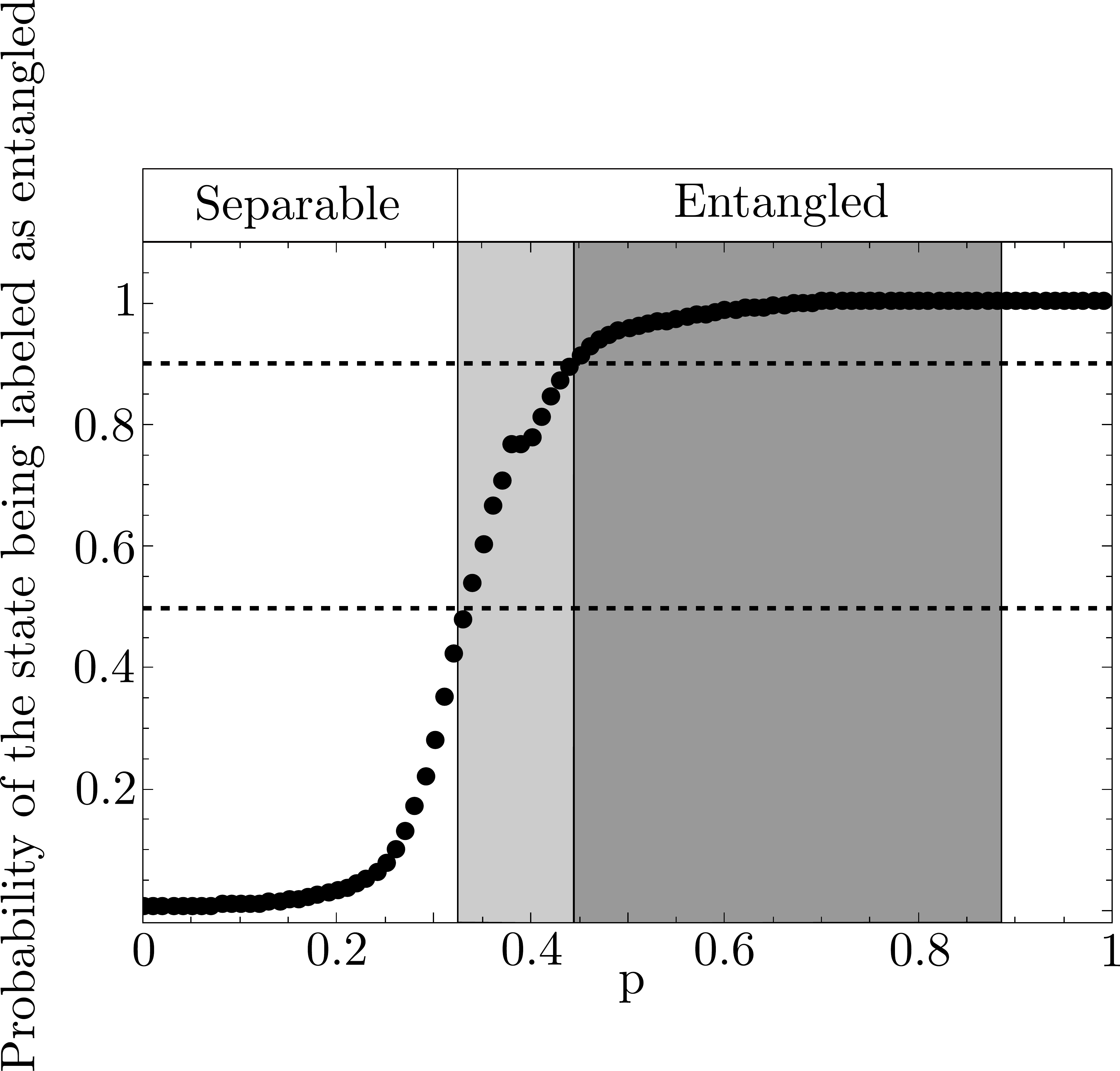}
		\caption{Results obtained by neural network and collectibility respectively from a real experimental data set $N=5$. Black full dots show the probability of a Werner state being labeled as entangled by the ANN. The light-gray area covers the values of $p$ which neither the neural network nor collectibility can classify correctly. The dark-gray area represents the range of $p$ values for which the ANN classifies the Werner states correctly and collectability fails. The dashed lines represent the decision thresholds $\epsilon = 0.9$ and $0.5$ respectively.}
		\label{fig:exp}		
		\end{center}
\end{figure}

\paragraph*{Experimental implementation.}To verify the network capability we decided to further test it on a set of real experimental data. For this purpose, we used the data set from the first-ever Collectibility measurement from 2016 \cite{PhysRevA.94.052334}. In that particular experiment, a class of Werner states of the form of $\hat{\rho}_w= p|\psi^-\rangle \langle \psi^-| + (1-p)\hat{\mathbb{1}}/4$, was investigated. $|\psi^-\rangle$ represents singlet Bell state, and $\hat{\mathbb{1}}/4$ stands for the maximally mixed state. We set the detection threshold to $\epsilon=0.9$ like in the previous comparisons of the neural network with collectibility, to be consistent and make test conditions as fair as possible. Results show that collectibility can classify states with $p>0.89$ as entangled witch corresponds with its theoretical prediction. The neural network, on the other hand, detects entangled states when $p>0.44$ (see Fig.\ref{fig:exp}). Note that it is known that Werner states are entangled for $p>\frac{1}{3}$.
\paragraph*{Conclusions.}We trained a neural network to classify general qubit states based on nonlinear collective witnesses. Our main goal was to compare the capability of this network against a prominent analytical representation of nonlinear witnesses: the  collectibility. The network can classify the general two-qubit states significantly more efficiently than collectability with $\text{Type-I error} < 1$ $\%$. The ANN also surpasses FEF, CHSH, and entropic witness when taught on 12 projections (the same amount needed by the mentioned analytical witnesses). Increasing the number of projection settings improves the ANN's decision even more. We further support this claim by using the network on a real experimental data set. The network confirmed its potential by correctly labeling a broad range of states where collectibility fails. Moreover, it achieved a $\text{Type-I error} = 0$ on Werner states. Our research  promotes the idea of using artificial intelligence towards a better understanding of the intriguing physical phenomena such as the entanglement. We have demonstrated that the neural network can quickly train to become a valid efficient collective entanglement witness.
\paragraph*{Acknowledgement.} Authors thank Cesnet for providing data management services. Authors acknowledge financial support by the Czech Science Foundation under the project No. 19-19002S. KB also acknowledges the financial support of the Polish National Science Center under grant No. DEC-2019/34/A/ST2/00081. JR also acknowledges internal Palacky University grant IGA-PrF-2020-007. The authors also acknowledge the project No. CZ.02.1.01./0.0/0.0/16\textunderscore 019/0000754 of the Ministry of Education, Youth and Sports of the Czech Republic.

\end{document}